\documentclass[11pt]{article}

\usepackage[utf8]{inputenc}
\usepackage{url}
\usepackage{graphicx}
\usepackage{subfigure}
\usepackage{multirow}
\usepackage{booktabs}
\usepackage{amssymb}
\usepackage{amsmath}
\usepackage{textcomp}

\title{\bf DIY Customized Speech Recognition for a Person with Speech Disabilities}

\author{
    {\bf Paulo A. Condado}\\
    \small CENSE -- Center for Environmental and Sustainability Research\\
    \small \& CHANGE - Global Change and Sustainability Institute\\
    \small NOVA School of Science and Technology, NOVA University Lisbon\\
    \small Campus de Caparica, 2829-516 Caparica\\
    \small{p.condado@fct.unl.pt}
    \and
    {\bf Fernando G. Lobo}\\
    \small CENSE -- Center for Environmental and Sustainability Research\\
    \small \& CHANGE - Global Change and Sustainability Institute\\
    \small DEEI-FCT, Universidade do Algarve\\
    \small Campus de Gambelas, 8005-139 Faro, Portugal\\
    \small flobo@ualg.pt
}
\date{}

\begin{document}
\tabcolsep=4pt
\maketitle

\begin{abstract}
General-purpose state-of-the-art automatic speech recognition (ASR) systems fail systematically for speakers with moderate to severe speech disabilities.
This paper describes the design, implementation, and evaluation of a customized ASR system built for the first author, a researcher with cerebral palsy and dysarthria.
We fine-tuned OpenAI's Whisper \emph{medium} and \emph{large-v3} models using Low-Rank Adaptation (LoRA) on audio samples recorded from the first author's own speech on consumer-grade hardware. 
On a held-out test set of 100 previously unseen utterances, the corresponding transcriptions resulted in a Word Error Rate drop from 49\% to 17\% for \emph{medium} and from 30\% to
12\% for \emph{large-v3}, while the Character Error Rate fell from 26\% to 7\%
for \emph{medium} and from 16\% to 5\% for \emph{large-v3}, 
indicating that the residual errors are minor phonemic substitutions. 
Based on these models, the first author developed
\emph{AVoz}, a mobile application that acts as a real-time
interpreter, converting his speech into synthesized voice for the
listener. All processing is performed locally, ensuring that no
audio data leaves the user's machine. We argue that the methodology
is replicable. Moreover, we publish the full LoRA configuration, the
implementation, and a step-by-step recording and training protocol,
so that other people with speech disabilities, or their family members
and caregivers, can build their own systems with a few hours of labeled  audio data and a couple of days of GPU time.
\end{abstract}

\section{Introduction}
\label{sec:intro}

This paper describes the development of an Automatic Speech 
Recognition (ASR) system customized to a person with cerebral palsy 
and speech disabilities. That person is the first author and lead 
developer of this work. The results obtained so far surpassed our initial expectations. 
Although customized to a single person, our work shows that the same methodology can be used 
by other people with speech disabilities to build their own customized 
ASR systems. 

ASR has been investigated for a long time, and significant progress 
has been made in recent years due to advances in artificial intelligence (AI), 
particularly with the development of deep learning and the Transformer 
architecture in artificial neural networks.
Among the most impressive state-of-the-art ASR systems is OpenAI's 
Whisper ~\cite{Radford:2023,Radford:github:2022}, a general-purpose, 
automatic speech recognition and speech translation system 
trained originally on 680,000 hours of multilingual audio data, and with the more recent \emph{large-v3} model trained on approximately 5 million hours of audio data. 
Although Whisper works very well for the majority of people, it performs poorly for people 
with speech disabilities. This is far from surprising because the 
training data consists largely of audio recorded from 
people with so-called ``normal" speech patterns. 

OpenAI released their Whisper models open source, making them readily 
available to the entire world. A variety of models are provided 
representing a trade-off between accuracy and computational performance. 
As an example, the \emph{small} model has 244 million parameters requiring 
approximately 2 GB of VRAM, but has a low accuracy. The \emph{large-v3} model 
contains 1.55 billion parameters and requires approximately 10 GB of 
VRAM. It is the highest accuracy model but is roughly 4 times slower 
than the small one.

We experimented Whisper with the \emph{medium} and \emph{large-v3} models, and 
tested the transcription of audio recorded from both authors of this 
paper. The first author, Paulo, has cerebral palsy from birth and has dysarthria. 
People unfamiliar with his speech, in general have 
trouble understanding him in an oral conversation. The second author, Fernando, does not have 
a speech disability.
Preliminary results immediately revealed that Whisper had trouble in 
properly transcribing Paulo's speech, regardless of the model used, 
making too many mistakes to be of practical use.  
With respect to Fernando, the accuracy was very good.

In this paper we show that what was considered nearly impossible some years ago 
is now within reach. By fine-tuning Whisper with 
a few hundred utterances of Paulo's own speech, we obtained
a transcription accuracy that, to our surprise, surpassed our
most optimistic expectations. The resulting models recognize Paulo's
speech with good accuracy and---crucially---do not merely memorize
the trained audio samples: they generalize to utterances they never
heard. This suggests that the models have in fact learned 
underlying patterns of Paulo's speech---the same patterns that
Paulo's family and close friends have internalized over years of
conversation.

Building on these models, the first author designed a mobile
application that allows him to communicate easily 
and effectively with anyone. The application acts as a real-time
interpreter. It recognizes Paulo's speech, transcribes it, and
converts it into synthesized speech, thereby conveying his message
to the listener in near real-time. In this paper we
describe both the fine-tuning process and the development of this
application.

The remainder of the paper is organized as follows.
Section~\ref{sec:related} reviews the literature on ASR for
atypical speech, parameter-efficient fine-tuning, and the
Do-It-Yourself (DIY)-assistive-technology movement.
Section~\ref{sec:motivation} describes Paulo's lived experience
and the 21-year journey that led to this work.
Section~\ref{sec:customization} details the customization of Whisper,
starting with its architecture, and the rationale for choosing Low-Rank Adaptation (LoRA)~\cite{Hu:2022} over full fine-tuning. It also details the data collection procedure and the training methodology.
Section~\ref{sec:results} reports the results obtained and discusses their relevance.
Based of those results, Section~\ref{sec:app} describes the development of \emph{AVoz}, a mobile
application that acts as a real-time interpreter.
Section~\ref{sec:ethics} discusses the ethical, privacy, and
socioeconomic implications.
Section~\ref{sec:future-work} presents future research directions, and
finally, Section~\ref{sec:conclusions} summarizes and draws the major conclusions of the work.


\section{Related Work}
\label{sec:related}

Automatic Speech Recognition has been investigated for a long time and significant progress has been made during the last few years
with state-of-the-art systems achieving an accuracy comparable to human-level performance. However, when used by people with moderate to severe speech disabilities, the accuracy degrades drastically making them impractical for everyday use. This is unfortunate as this group of individuals are the people who could benefit the most from the technology. 

\subsection{ASR for dysarthric speech}

Recognition performance depends on several factors,
namely speech intelligibility, the amount of speaker-specific training data,
and the adaptation strategy~\cite{Rudzicz:2007,Christensen:2012,Mustafa:2014}.
Hawley et al.~\cite{Hawley:2007} showed that small-vocabulary
speaker dependent systems can outperform speaker independent ones for speakers
with severe dysarthria. 

The use of deep learning techniques has transformed the
field~\cite{Prabhavalkar:2023,Qian:2023}. Recent approaches use
self-supervised and cross-lingual speech representations~\cite{Hernandez:2022,Hu:2023},
as well as new datasets, such as the one collected by the \emph{Speech Accessibility
Project}~\cite{HasegawaJohnson:2024}, to mitigate the lack of training data.
Moreover, recent studies also address continuous and conversational speech 
in real-world scenarios~\cite{Tobin:2024,Jiang:2024}. An interesting study was presented by Koenecke et al.~\cite{Koenecke:2024}, who showed that Whisper occasionally \emph{hallucinates} entire phrases that were never spoken, and 
that this happens more often for speakers with aphasia. It is important to 
stress that aphasia is a language disorder rather than a motor speech disorder, 
and therefore poses different challenges. Nevertheless, this study shows that Word Error Rate (WER)
alone is not enough to evaluate whether an ASR system is useful for a person
with a speech disorder.

The work most closely related to ours is \emph{Project
Euphonia}~\cite{Shor:2019} that showed that
\emph{customized} models, fine-tuned with recordings of a single speaker, can
substantially reduce WER for dysarthric and accented speech. Green
et al.~\cite{Green:2021} went further and showed that, in some cases, these
models outperform human listeners on short phrases, and Tobin and
Tomanek~\cite{Tobin:2022} showed that useful gains can be obtained with small
amounts of speaker data. More recently, Raja et al.~\cite{Raja:2025}
fine-tuned Whisper to compare \emph{idiosyncratic} (speaker-specific) and
\emph{normative} (multi-speaker) models for dysarthric speakers, including
speakers with cerebral palsy, and Project Euphonia released an open-source
toolkit, which includes a mobile application to record phrases and notebooks
to fine-tune an open model for a specific speaker~\cite{Euphonia:2026}.
Therefore, customizing an ASR model for a single speaker is not new in
itself. However, in all these studies the customization is carried out by
research teams, mostly for the English language,
and the person with a speech disorder
participates as a \emph{data contributor}, not as the designer and builder
of their own recognizer~\cite{Shor:2019,Green:2021,Tobin:2022}. 
Even the Euphonia toolkit is aimed at developers
rather than at the people who will use the resulting models, and, by
default, it relies on cloud services, namely Google Colab for training and
Google Cloud Run for transcription~\cite{Euphonia:2026}. It is important to
stress that speech recordings are sensitive personal data, especially when
they reveal a health condition~\cite{Condado:2023}. In our approach, both the
fine-tuning and the recognition are performed locally, on hardware
controlled by the user, so that the recordings never leave their devices.
To the best of our knowledge, this is the first study in which a researcher
with speech disability conceived, executed, and evaluated the customization,
and documented the process in a form that another user can replicate without
the help of a research team. As we shall discuss in 
Sections~\ref{subsec:r-diy} and~\ref{subsec:m-maker}, this work also 
follows a constructionist approach~\cite{Papert:1993}, in which the 
first author designed and built a 
solution to an issue faced by himself and others in a similar condition~\cite{Condado:2026}.

Speech recognition is only one component of a broader set of language
technologies for people with complex communication needs. In the specific area of augmentative and alternative communication (AAC),
systems based on large language models (LLMs) have recently been explored to
support users with aphasia~\cite{Zeng:2024}, and Valencia
et al.~\cite{Valencia:2023} studied how LLMs can either enhance or impede
communication for AAC users. Other work has
explored eye-gaze and image-recognition conversational support for people
with amyotrophic lateral sclerosis~\cite{Kane:2017} and word prediction for AAC users~\cite{Garcia:2014}.

\subsection{Fine-tuning and parameter-efficient adaptation of ASR models}
\label{subsec:r-peft}

Classical \emph{speaker-adaptive} ASR relies on techniques such as maximum a
posteriori (MAP) estimation~\cite{Gauvain:1994} and maximum likelihood
linear regression (MLLR)~\cite{Leggetter:1995}, or simply on training
speaker-dependent acoustic models. These techniques are also the basis of
much of the early work on dysarthric ASR~\cite{Christensen:2012,Mustafa:2014}.
With the rise of large pre-trained models, such as
wav2vec~2.0~\cite{Baevski:2020}, HuBERT~\cite{Hsu:2021}, and
Whisper~\cite{Radford:2023}, the ASR community shifted to \emph{fine-tuning}.

We refer to fine-tuning as the process of taking a pre-trained model and training it further on a smaller, specific dataset, so it can perform a particular task very well.  
The two major strategies commonly used are to either use \emph{full fine-tuning} or a 
\emph{parameter-efficient fine-tuning} (PEFT) method. 
Full fine-tuning consists of updating all model parameters of a pre-trained model based on the additional training data. It is a powerful method but for large models requires a significant amount of computational power and memory, making the overall training process slow and expensive. Moreover it is prone to
overfitting when only a few hours of speech is available, which is precisely the case for an individual speaker.

Fortunately, PEFT methods address
both problems. Instead of updating all the weights of the pre-trained model,
these methods keep them frozen and only train a small number of additional
parameters, namely adapters~\cite{Houlsby:2019} or low-rank matrices, as is the case of LoRA~\cite{Hu:2022}. 
LoRA is motivated by the observation that fine-tuning
pre-trained models has a low ``intrinsic dimension''~\cite{Aghajanyan:2021},
and QLoRA further reduces memory requirements by fine-tuning on top of a
quantized model~\cite{Dettmers:2023}. In practice, this means that a model such as Whisper can be fine-tuned on a personal computer equipped with an inexpensive consumer-grade graphics processing unit (GPU) or, ideally, a computer powered by Apple silicon. PEFT
methods have been used to extend Whisper to new languages~\cite{Song:2024}, to
adapt it to child speech~\cite{Liu:2024}, and to adapt pre-trained models to
individual dysarthric speakers~\cite{Qi:2023}.

In these works, the adaptation target is typically a \emph{language}, an
\emph{accent}, or a \emph{population} of speakers, and the adaptation is
designed and carried out by ASR researchers. Our work differs in that the
adaptation target is \emph{a single human being} --- Paulo --- and the
adaptation is built \emph{by} that person with the collaboration of the second author.

\subsection{DIY assistive technology and first-person research}
\label{subsec:r-diy}
Our approach is inspired by Seymour Papert's \emph{constructionism}, namely
the idea that people learn best when they are actively \emph{building}
something that is meaningful to them and that can be shared with
others~\cite{Papert:1993,Stager:2016}.
In \emph{Mindstorms}, Papert contrasted two ways of using computers in
education: one in which ``the computer is being used to program the child''
and another in which ``the child programs the computer''~\cite{Papert:1993}.
We argue that DIY assistive technologies follow the same principle, with a
person with disabilities in the place of the learner. First, the person stops
being a passive \emph{user} of technology designed by others and becomes its
\emph{builder}. Second, the solution is personally meaningful, since it
addresses the builder's own needs. Third, the act of building is itself a
source of learning and autonomy. Papert's
ideas apply directly to our work. A general-purpose ASR system
``programs'' users with speech disorders, forcing them to slow down, to
repeat themselves, and to adapt to the system. On the other hand, in a DIY
approach, it is each person who ``programs'' the recognizer, adapting it to
their own speech. Moreover, Papert advocated programming environments with a
``low floor'', allowing children to start building with little
effort~\cite{Resnick:2009}. In the same way, open pre-trained models and PEFT
methods have considerably lowered the floor for building a customized ASR
system. In other words, it is no longer necessary to have a large research team or
an expensive infrastructure. Anyone with the appropriate hardware, namely a
personal computer equipped with a consumer graphics card or a computer powered
by Apple silicon, should be able to replicate the process described in this
paper and build an ASR model adapted to their own speech.

These ideas have influenced assistive technology since at least the early
2010s. As argued by Hurst and Tobias~\cite{Hurst:2011}, DIY technology can
empower individuals with disabilities to create and adapt their own
assistive devices. Subsequent studies addressed the challenges of DIY
assistive technology in practice~\cite{Hook:2014}, the sharing of assistive
designs in online maker communities~\cite{Buehler:2015}, the role of
community-based and volunteer fabricators, such as
e-NABLE~\cite{Parry-Hill:2017}, and the emergence of public DIY assistive
technology programs~\cite{Higgins:2024}. Closer to our setting, Thorsen et
al.~\cite{Thorsen:2023} described a workflow in which people with cerebral
palsy co-create assistive devices using 3D printing. In all these cases, the designs are shared so that other people
can replicate and adapt them to their own needs. However, most of these
studies focus on \emph{physical} artifacts. A closer analogy
to our work is the ``DIY closed-loop'' (\#WeAreNotWaiting) movement in health
technology~\cite{Lewis:2019}, in which people with diabetes build and
run their own automated insulin delivery systems using open-source software.
As in our case, the person who lives with the condition is also the one who
configures, tests, and maintains a data-driven system tuned to their own body.

Our work is also related to first-person and autobiographical approaches in
HCI, in which researchers design for and study their own
lives~\cite{Neustaedter:2012,Desjardins:2018,Lucero:2019}, as well as to
disability-led research, which values the lived experience of disabled
researchers~\cite{Hofmann:2020}. For instance, Mo et al.~\cite{Mo:2025}
recently reported their experience as a PhD student with cerebral palsy, 
and so did ourselves reporting our experience in building assistive technologies~\cite{Condado:2026}.

\subsection{Positioning of this work}
\label{subsec:r-position}

In summary, the literature shows that: (i) general-purpose ASR systems still
perform poorly for people with moderate to severe speech disabilities; (ii)
customizing a model for a single speaker can close much of this gap, but so
far it has been carried out by large research teams, with the speaker as a data
contributor; (iii) PEFT methods, such as LoRA and QLoRA, allow large open
models to be adapted on consumer-grade hardware; and (iv) DIY approaches have
shown the value of people with disabilities building their own assistive
technologies, but mostly for physical devices and, with the exception of DIY
closed-loop systems for diabetes, rarely for data-driven systems. Our work
brings these four lines of research together.

\section{Motivation and Lived Experience}
\label{sec:motivation}

Being able to automatically recognize Paulo's speech has been a long-term 
goal of ours for almost 21 years. But it was such a difficult task that we 
almost gave up on it, and for many years we set that goal aside as it 
seemed nearly impossible to attain. 

Most people unfamiliar with Paulo's speech have a lot of trouble 
understanding what he says. In contrast, for those accustomed to talk with him---like Fernando, the second author of this paper and his long-time research collaborator---understanding Paulo is not a problem.

Throughout the years, Fernando got used to Paulo's speech and nowadays 
understands him almost perfectly, even though occasionally he still asks 
him to repeat himself. Similarly, Paulo's family members have no problem 
understanding him because they are completely used to his way of speaking. 
This observation, which was already made in Paulo's PhD dissertation~\cite{Condado:2009}, reveals that there are patterns in Paulo's 
speech that people are able to learn if they spend enough time interacting 
with him. And if people can learn those patterns, a machine can potentially 
also learn them---after all, learning patterns from data is what Machine 
Learning (ML) is all about.

\subsection{How Paulo communicates}
\label{subsec:how-Paulo-communicates}
Paulo is a very social person and has always enjoyed meeting and talking 
with other people. Unfortunately, only those accustomed to interact with him 
are able to easily communicate orally with him. Because of that, Paulo often resorts 
to the utilization of speech synthesis, also commonly referred to as 
text-to-speech (TTS), whenever he needs to say something of moderate length. In such occasions,
he types beforehand what he wants to say and uses the TTS system to deliver the speech.

Although not perfect, this method has allowed Paulo to conduct public presentations.
With respect to writing, Paulo never  had a real problem other than typing slowly due to the hand-motor coordination  problems that he has as a result of his cerebral palsy condition. 
Overall, for those not used to Paulo's way of speaking, written 
communication tends to be the most effective.

Paulo could use a TTS system during a regular conversation, but he 
rarely does so because the flow of the conversation becomes awkward: it takes time for Paulo to 
type what he wants to say and let the TTS system do the rest. 
Most people simply do not have the time and patience to wait that amount of time, and 
so Paulo just talks in his own way hoping the other person will understand 
him. What usually happens is that the other person does not understand (or
understands very little), and the conversation quickly comes to a halt.

Although both are speaking the same language, it is as if Paulo is speaking 
a foreign language that the other person does not understand. In fact, only 
those accustomed to Paulo understand that ``foreign language'' --- as is the 
case of Fernando and Paulo's close friends and family. When Paulo is talking 
with someone else and the other person does not understand him, if Fernando 
is nearby he often acts as an interpreter, repeating what Paulo just said, 
precisely as if Paulo was speaking in a foreign language.

Indeed, Fernando recalls the first interactions he had with Paulo more than 
25 years ago, when Paulo attended a database course at the University of 
Algarve that he was teaching at the time. Paulo often raised his hand 
to ask questions, but Fernando could not understand what he was saying, and so, one 
of Paulo's classmates who was already familiar with his speech, would take 
the role of the interpreter. As time went by, Fernando started to get used 
to Paulo's speech and soon there was no need for an interpreter anymore.

For Paulo it has always been frustrating not being able to be understood. 
Using TTS technology mitigates part of the problem but not completely, 
especially in a real-time conversation because Paulo is a slow typer.
Another important aspect worth mentioning is that Paulo prefers to speak with 
his own voice rather than with an artificial voice generated by TTS, and that's understandable---our voice is something personal that resonates with one's inner self. This observation is consistent with recent research work by Weinberg et al.~\cite{Weinberg:2026} who refer that ``\emph{Voice is a central element of identity.}"

Paulo also observes that although some make an effort to understand him, the majority of people do not, and quickly give up during a conversation. Perhaps this is a sign of the busy 
life that everyone seems to live: very few people have patience and take 
the time to listen to him carefully.

\subsection{Paulo, the maker}
\label{subsec:m-maker}
Paulo has been a technology enthusiast from a very early age, having 
learned the basics of computer programming during his childhood. That 
exposure made him believe that he could be a maker, and he became an adept 
of the DIY approach. A recent experience report published at the 2026 ACM 
ASSETS conference presented an auto-ethnographic study of his life~\cite{Condado:2026} 
and describes, among other things, various technological solutions that he 
designed to address his own needs~\cite{Condado:2008,Condado:2015,Condado:2022,Condado:2023}. 

One such solution that is particularly relevant to the present work was 
the \emph{EasyVoice} program~\cite{Condado:2007,Condado:2008} that Paulo 
designed and built. 
EasyVoice integrated text-to-speech with Voice-over-IP (VoIP) to allow a person with speech
disabilities to make a phone call and 
type the intended messages, which were subsequently converted to audio and
sent to the other person. EasyVoice was tested with Skype, which at the 
time was the most popular IP-based voice-call application and provided 
an Application Programming Interface (API) for developers. 

With EasyVoice, Paulo was able to establish phone calls with other people 
anywhere in the  world. A communication barrier had been mitigated, but 
not entirely. Although EasyVoice had features to accelerate text input, such 
as word completion, it still required a substantial amount of time for Paulo 
to  type what he wanted to say, making the flow of the conversation a bit awkward and inefficient. Notwithstanding, 
the system at least made voice communication possible.

At the time, we immediately recognized that what would be ideal was to have a speech 
recognizer that could transcribe Paulo's speech to text and feed the resulting 
text into the TTS system, eliminating the time required to type messages and 
making the conversation flow smoother. Such a system would transform an 
indiscernible speech into a discernible one. Unfortunately, at the time ASR 
systems did not produce accurate enough transcriptions for people with ``normal'' speech patterns, let alone someone with speech disorders.
Fast forward 21 years, and current state-of-the-art ASR systems 
are producing impressive results, being able to recognize most people’s 
speech very well. The accuracy, however, deteriorates 
quickly for people with speech disabilities.

Our ASSETS 2026 experience report~\cite{Condado:2026}  concludes with the idea that people 
with disabilities should be encouraged to be makers and try to build 
artifacts that are meaningful to them, playing an
active role in the design, development, and evaluation of the technologies that they use every day.
Indeed, the present work can be 
seen as the latest example of our own advice. Given that current ASR systems
perform poorly for Paulo’s speech, he embarked on customizing one for his own needs.

Our goal is not only to fine-tune an ASR to recognize Paulo's speech, but also to
describe the whole process in a way that allows other people with speech
disorders, or their family members and caregivers, to replicate it with
inexpensive and widely available hardware. 
The present work also describes the development of a mobile application that
uses the fine-tuned model to recognize Paulo's speech, transcribe it, and
convert it into synthesized speech, in order to ``translate'' him to other
people.


\section{Customizing Whisper with LoRA}
\label{sec:customization}

Whisper~\cite{Radford:2023,Radford:github:2022} is an end-to-end speech
recognition and translation model based on the encoder--decoder Transformer
architecture~\cite{Vaswani:2017}. The original models, such as \emph{small}
and \emph{medium}, were trained on 680,000 hours of multilingual audio,
while \emph{large-v3} was trained on \emph{1 million hours of weakly labeled audio
and 4 million hours of pseudo-labeled audio}~\footnote{\url{http://github.com/openai/whisper/discussions/1762/} (Accessed on Oct 12, 2026).}. 
Unlike traditional ASR systems, which require separate components, namely
feature extraction, acoustic, pronunciation, and language models, end-to-end
systems use a single neural network to map audio directly to text.
The success of deep learning together with more generic model architectures has lead 
end-to-end models to become the prominent ASR approach.

\subsection{Initial testing and the need for customization}
\label{subsec:initial-testing}

With the initial help provided by the Whisper github
repository~\cite{Radford:github:2022}, Paulo wrote a small Python
program to test Whisper's ability to transcribe his speech. The results
were very poor regardless of the model used, with a large number of transcription errors. When tested
with Fernando, who does not have a speech disorder, Whisper's
transcriptions were very good. Paulo also evaluated other freely
available ASR systems, including Vosk\footnote{%
  \url{https://github.com/alphacep/vosk-api} (Accessed on Oct 12, 2026).} and
sherpa-onnx\footnote{\url{https://github.com/k2-fsa/sherpa-onnx} (Accessed on Oct 12, 2026)}, but the results were similarly poor. 

These results confirm what we already suspected: recognizing the speech
of a person with a speech disorder is a much harder task, and presents
an important challenge even for modern state-of-the-art ASR systems.
Although Whisper has been trained on a large amount of diverse audio
data, that data is unlikely to contain much audio (if any at all) of people with speech
disorders, and so Paulo's speech is not well represented in the
training distribution. This is a typical \emph{out-of-distribution}
scenario in machine learning, and the natural solution is to provide the
model with additional speaker-specific examples.

\subsection{Data collection}
\label{subsec:data-collection}

Paulo started to read aloud a variety of pre-written sentences in European Portuguese, his mother tongue, 
and recorded the corresponding audio. 
Simultaneously, we looked at ways of fine-tuning Whisper for the additional training data.

The recording setup was as follows. All recordings were made in
Paulo's office, using the built-in microphone of an HP webcam
positioned approximately 50 cm to 1 m in front of him. The computer
was running Ubuntu Linux. Audio was captured directly from the ALSA
default input device using the following command:
\begin{center}
\texttt{ffmpeg -f alsa -i default -ac 1 -ar 16000 -c:a pcm\_s16le t0100.wav}
\end{center}
which produces a mono, 16 kHz, 16-bit PCM WAV file --- the same format
and sample rate that Whisper expects as input.
Recordings were made between the end of August and the beginning of
September 2026. 

The average duration of an audio sample was approximately 
16 seconds, and the final dataset used for training and validation consisted of 765 audio samples with a total duration of 3 hours and  22 minutes. No pre-processing of any kind was applied to
the audio. The recordings are exactly what the microphone
captured, and the model was trained on that raw signal.

The writing of the sentences and the collection of the audio
recordings were handled as follows. Paulo wrote and recorded 
all the training and validation sentences.
Since he was the author of the original text, the sentence itself served as its reference
transcription.

To create the test dataset, Fernando wrote a separate set of 100 sentences, which
was reviewed by Paulo to ensure that they were not included (by coincidence) in 
the training and validation datasets. 
This was a deliberate 
methodological choice. Since Fernando did not know the content of Paulo's 
training and validation datasets, this procedure ensured that the test 
sentences were novel to the model, allowing us to evaluate its performance 
on previously unseen data. 

The test sentences were recorded on a different day from the training 
and validation sentences to reduce the influence of session level variations, 
such as fatigue or changes in recording conditions. This decision also 
enables us to assess the model's performance on recordings collected in 
a separate session.

\subsection{LoRA, model selection, and training methodology}
\label{subsec:lora-method}

We chose Low-Rank Adaptation (LoRA)~\cite{Hu:2022} to customize Whisper for Paulo's speech, and we did so instead of a full fine-tuning strategy due to reasons discussed earlier in Section~\ref{subsec:r-peft}, namely: lower memory
requirements, faster training, and no additional latency at inference time.
LoRA keeps the weights of the pre-trained model frozen and only adds to it a small number of additional trainable parameters corresponding to two low-ranked matrices into each layer of the Transformer architecture, dramatically lowering the computational effort.
It is a simple and elegant approach that has been shown to work quite well in several applications that needed to adapt to particular domains. 

We opted to fine-tune the Whisper \emph{medium} and the \emph{large-v3} models.
This decision was driven by practical considerations.
\emph{Medium} offers a favorable trade-off between accuracy and memory,
making it feasible to fine-tune on a personal computer equipped with a
single consumer-grade GPU or on a computer powered by
Apple silicon. \emph{Large-v3} was retained as a reference point to
assess how much accuracy is lost by using the \emph{medium} model instead.

\subsubsection{Evaluation metrics}

ASR systems are typically evaluated using the Word Error Rate (WER), defined
as $(S+D+I)/N$, where $S$, $D$, and $I$ are the numbers of word
substitutions, deletions, and insertions, and $N$ is the number of words in
the reference transcription~\cite{Morris:2004}. 
We note however 
that WER has well-known limitations. First, all errors have the same weight,
regardless of their impact on the meaning of the
sentence~\cite{Morris:2004}. Second, since insertions are counted
and $N$ depends only on the reference, WER has no upper bound and can exceed
100\%~\cite{Morris:2004}. This may happen, for instance, when the model
\emph{hallucinates} words that were never spoken~\cite{Koenecke:2024}.
Whisper, as well as most ASR systems, reports WER and complements it with the Character Error
Rate (CER), computed in the same way at the character level. 
We will also be using these two metrics in our experiments.

\subsubsection{Fine-tuning configuration}

The dataset of 765 audio samples and corresponding text references was split into 85\% for training and 15\% for validation. 
All fine-tuning runs were performed using the Hugging Face
Transformers~\cite{Wolf:2020} and {PEFT~\cite{PEFT:2022} libraries. 
The pre-trained Whisper model was kept frozen in all runs, and
only the LoRA matrices were updated. Following the configuration recommended
in the original LoRA paper~\cite{Hu:2022}, the adapters were applied to the
query and value projections (q\_proj and v\_proj) of all
attention layers of the encoder and decoder.

The models were trained with the AdamW optimizer~\cite{Loshchilov:2017}, a weight decay of 0.01, 5 warmup steps followed by a linear learning rate decay, and full precision. Due to memory constraints, we used a batch size of 2 with 8
gradient accumulation steps, which corresponds to an effective batch size of 16. Each model was trained for a maximum of 13 epochs, and the checkpoint
with the lowest WER on the validation set was selected, which usually
occurred around epoch 10. 
These values were defined in an initial
exploratory phase, in which we also trained models for 3, 5, and 8 epochs.

The rank $r$ and the learning rate $\eta$ were tuned based on the WER validation performance obtained from experiments where we varied $r \in \{2,4,8,16\}$ and $\eta = k \times 10^{-5}$ for $k \in \{3,4,5\}$.
When different ranks achieved comparable validation performance, we preferred the smaller rank to reduce the number of trainable parameters and the associated computational overhead, as well as lower the risk of overfitting. This follows the empirical rank-selection motivation investigated by~\cite{Hu:2022}, while allowing a suitable rank to be determined for our specific task.
We use a scaling factor of $\alpha/r=2$, and a dropout of 0.05.

Based on the results of the hyperparameter sweep experiments, we settled on rank $r=8$ for both models, and a learning rate $\eta = 5 \times 10^{-5}$ for \emph{medium} and 
$\eta = 4 \times 10^{-5}$ for \emph{large-v3}.

All experiments were conducted on a Mac Studio equipped with an Apple M4 Max chip and 64 GB of unified memory. This configuration provides sufficient computational resources to run and fine-tune speech recognition models locally, without requiring dedicated high-end GPUs or cloud-based computing services. In recent months, there has also been a growing interest in using Apple computers, particularly the Mac mini, to run LLMs locally. In fact, the increasing demand for these computers illustrates the interest in using consumer-grade hardware to run AI models locally, rather than relying on cloud-based services~\footnote{The growing demand for the Mac mini for local AI workloads has also been reported by Larsen~\cite{Larsen:2026}.}.

The hold-out test set was never used to make any decisions during fine-tuning, so that a true test on unseen data can be properly assessed.
We publish the code of our implementation and a  step-by-step training protocol, in the spirit of the DIY assistive technology movement~\footnote{Available at  \url{https://github.com/pacondado/ASR-for-people-with-speech-disabilities}}.

\section{Results}
\label{sec:results}

Earlier in Section~\ref{subsec:initial-testing} we refer that we did preliminary tests with Whisper and observed that its accuracy with Paulo's speech was poor. That initial exploration was rather crude, but enough to conclude that customization was needed. Table~\ref{tab:results} reports the results in a more precise, quantitative manner, using the WER and CER metrics. We do so for the original baseline models as well as for their fine-tuned versions.

\begin{table}[h]
	\centering
	\small
	\caption{Accuracy of the Whisper \emph{medium} and \emph{large-v3} models, with and without LoRA fine-tuning,  on the hold-out test set of  $n=100$ sentences spoken by Paulo.}
	\label{tab:results}
		\begin{tabular}{lcc}
         \toprule
		\textbf{Model} & WER (\%) & CER (\%) \\
		\midrule
		\emph{medium}                                            & 49.2 & 26.0 \\
		\emph{medium}  fine-tuned with LoRA   & 17.0 & 7.3 \\
		\emph{large-v3}                                           & 30.4 & 15.6  \\
		\emph{large-v3} fine-tuned with LoRA   & 12.4 & 5.4 \\
		\bottomrule
	\end{tabular}
\end{table}

The overall picture is clear. Fine-tuning Whisper with LoRA on as
few as 765 sentences of Paulo's own speech reduces the WER and CER metrics substantially. 
For \emph{medium}, the WER drops from 49.2\% to 17.0\%. 
For \emph{large-v3}, the WER drops from 30.4\% to 12.4\% . 

These numbers are not merely statistical. 
A conversation built on a WER of around 30\% 
requires the listener to constantly guess, ask for repetition, and
lose patience. A WER of 12.4\% means that approximately 7 out of every 8 words are correct, and the conversation can flow. For Paulo,
whose primary complaint during his entire life has been that
people stop trying to understand him, this difference is not a
marginal improvement. It is the difference between a system that
frustrates and a system that works.

The fine-tuned \emph{large-v3} model outperforms \emph{medium}, but the \emph{medium} is not that bad either.
For a person with speech disorders who wants to
develop a customized ASR system on a consumer-grade computer (e.g.,
Mac mini), this is a practical and encouraging result.
The CER obtained with the fine-tuned \emph{large-v3} model was only around 5\%
This means that, at the character level, the
transcription is substantially more accurate than the word-level
metric suggests. This is exactly what we expected for dysarthric
speech. A single character error in a word often results in a minor phonemic error, such as producing /d/
instead of /t/, and counts as a full WER
error, even though the semantic content is largely preserved. The
CER reveals that the model has in fact learned the phonetic
structure of Paulo's speech, and that many of the residual WER
errors are minor phonemic differences rather than wholesale misrecognitions.

\subsection{Discussion}
\label{subsec:results-discussion}

Shor et al.~\cite{Shor:2019} showed that customized models can substantially
reduce WER for dysarthric and accented speech, but their experiments were
conducted for the English language. Green et al.~\cite{Green:2021}
showed that, in some cases, customized models outperform human listeners on
short phrases, again in English and on a curated dataset. More recently,
Raja et al.~\cite{Raja:2025} fine-tuned Whisper to compare idiosyncratic and
normative models for dysarthric speakers, including speakers with cerebral
palsy, also in English, and Project Euphonia released an open-source toolkit
to fine-tune an open model for a specific speaker, which, by default, relies
on cloud services~\cite{Euphonia:2026}. In all of these studies, the
customization was designed and executed by large teams of researchers or developers, and
people with speech disorders participated as data contributors. In our work,
the customization was conceived, conducted, and evaluated by the first
author, a researcher with a speech disorder, in European Portuguese, on a
single consumer-grade machine, and without sending any data to external
servers. We believe this is an important distinction, and one that is
directly relevant to the accessibility and sustainability of assistive
technologies.

The test sentences were written by Fernando, who deliberately did not have access
to Paulo's training and validation datasets. The model had never heard
these sentences during training, and yet it recognized them with a
WER of 12.4\% and a CER of 5.4\%. This is a strong evidence that the model learned relevant underlying
patterns of Paulo's speech, akin to the patterns that Paulo's family
and close friends have internalized over years of
conversation~\cite{Condado:2026}.

A WER of 12.4\% and a CER of 5.4\% mean that 7 out of every 8 words are transcribed correctly, and that at the character level there are only a few mistakes once in a while.
These results are sufficient
for developing a mobile application to act as a reliable
interpreter in everyday conversations. The residual errors, while
still present, are minor phonemic substitutions that a human
listener can resolve from context. This is, in our experience, the
level of accuracy at which Paulo can hold a conversation without
needing to repeat himself after every sentence, a significant qualitative
change from the Whisper baseline models.

\section{Development of a Mobile Application}
\label{sec:app}

The results reported in the previous section show that the
fine-tuned models recognize Paulo's speech with sufficient accuracy
for practical use. 

Speech recognition can be used for a variety of applications, such as controlling devices by giving simple voice commands, dictation to allow writing text, and real-time subtitling, to name a few. But for someone like Paulo, the most important application is to be able to have an oral conversation with someone else and be understood. While that is something that most people take for granted, the same cannot be said for someone with a speech disability. As a matter of fact, being able to be understood in a oral conversation was the driving force that led us into this work. 

We aim for an application that can record an audio sample of Paulo speaking, and then perform speech-to-text using the customized ASR, followed by text-to-speech, effectively transforming an indiscernible speech into a discernible one.

An important aspect worth mentioning is that many errors captured by the WER measure are irrelevant in the context of such an application. For example, if Paulo speaks ``\emph{I want ice cream}" and the ASR system transcribes it to ``\emph{I want I scream}", technically there are 2 word errors, but if the resulting text is fed to a TTS system such errors are undetected because ``\emph{ice cream}" sounds exactly the same as ``\emph{I scream}". Indeed, this is an example of what linguistics call an \emph{oronym}:  a sequence of words that sounds identical or nearly identical to a different sequence of words when spoken.

Before designing a dedicated application, 
the first author wanted to confirm that this level of accuracy was 
maintained in a real conversational environment rather than in a 
controlled setting.
He therefore wrote a small Python program using the Flet
framework~\footnote{Available at \url{https://flet.dev/} (Accessed on Oct 12, 2026)}, which he ran directly on his own computer. The program
listens to the microphone in real-time, sends the audio to the
fine-tuned model running locally, displays the transcription, and
plays back the result through a TTS engine. Preliminary testing in 
spontaneous conversation suggest that the accuracy observed on the 
held-out set is not an artifact of the read-aloud conditions, although 
a controlled comparison remains to be conducted.

Subsequently, the first author designed a mobile application called
\emph{AVoz}\footnote{The name \emph{AVoz} alludes to several things, and is also an oronym.
	 It has the same sound as ``\emph{A voz}", the same sound as ``\emph{A vós}", and the same sound as ``\emph{Avós}", which mean ``\emph{The voice}", ``\emph{To you}", and ``\emph{Grandparents}", respectively, in the Portuguese language.  
	The letter 'A' in 'AVoz' can also stand for  \emph{Assistida},  making the term allude to \emph{Assistive Voice}.}, 
drawing on his prior experience in designing assistive technologies.
The same design principles that have been guiding our work were applied here as well: a
researcher with disabilities identifies the problem, sketches
the interface, writes the code, tests the solution in his own daily 
life, and decides when it is good enough for his own use, while
simultaneously initiating a process that may lead to solutions
for other people with similar needs through an iterative process
of development. This approach follows a
user-centered design philosophy, in which the needs and
experiences of the person who will use the technology guide
its development and evaluation~\cite{Condado:2026}.

\subsection{Architecture}

For practical reasons, and in particular because of the
computational limits of current cost-effective mobile devices, we 
opted for an extremely simple architecture. The application
presents just one large button.
When the user presses the button,
the application starts capturing audio from the device's microphone. When
the user presses the button again, the recording stops, and the
audio is sent over an encrypted channel to a server running on
the Mac Studio, where the fine-tuned Whisper model performs
inference. The resulting transcription is returned to the
application over an encrypted channel, and the application
immediately converts the text into synthesized speech, which is
played through the phone's speaker. The entire cycle, from the
moment the user finishes speaking to the moment the listener
hears the synthesized voice, is extremely fast and enables a user
with speech disabilities to have a real-time conversation. In our 
measurements, on a local area network, the process takes on 
average 1.85 seconds.

The server address is not
hard-coded in the application. The user can change it at any
time, making its configuration flexible. The application can
connect to a server at home, at the university, or anywhere else, 
depending on the user's location and
the machine available to run the model while preserving their
privacy.

\subsection{Privacy by design}

The architecture described above is not
merely a matter of convenience. It is a deliberate decision, based on our
privacy and data protection concerns. All inference is performed on a
computer that the user owns and controls. The audio recorded by the mobile
application is sent to this computer over an encrypted channel
 and is never stored on third-party servers. This is
consistent with the principles discussed in our previous work on security
and privacy in assisted living environments, namely that data collected
about users' daily activities should not be sent to external servers, but
stored on local servers to protect their
privacy~\cite{Condado:2022,Condado:2023}. 

\subsection{A functional prototype, not a finished product}

The application was built following the same
iterative, do-it-yourself process that produced EasyVoice 
and other assistive solutions~\cite{Condado:2008,Condado:2015,Condado:2022,Condado:2023}. 
The first author identified the need, designed the
interface, wrote the code, tested it in his own daily life, and
refined it over several weeks. Today it is absolutely functional
and is used in everyday situations, from ordering a coffee to
explaining a research idea to a collaborator who does not
understand his speech.

Nonetheless, \emph{AVoz} should be viewed as a
\emph{functional prototype} rather than a finished product. This is
consistent with the trajectory of our previous work~\cite{Condado:2008,Godinho:2015,Condado:2015,Condado:2022}. We intend to follow the same methodology
here. In line with a user-centered design
approach~\cite{Condado:2026}, we plan to evaluate both the
application and the fine-tuning process with other people with
speech disorders, so that the design reflects a broader range of needs,
speech patterns, and daily contexts. The model will be fine-tuned for
each individual user, and the application is therefore expected to
evolve iteratively in response to their feedback. The goal is not to
present a one-size-fits-all solution, but to demonstrate that the
methodology is replicable and to learn from other users' experiences
how the system should adapt.

\subsection{First impressions}

Paulo has been using \emph{AVoz} in a spontaneous way and it works remarkably well~\footnote{A video demonstrating the mobile application in use by Paulo is available at  \url{https://youtu.be/nKMi30XIkx4}}, ``translating" his speech into an equivalent TTS audio, and allowing him to communicate orally with others without needing to have someone familiar close to him to act as a human ``interpreter". \emph{AVoz} still makes a few little mistakes once in a while, but nothing severe, and most of the time the other person is able to pick up the intended meaning from the context. Moreover, Paulo also hears the TTS output and he's able to repeat himself when a mistake occurs. This behavior is not that different from what happens when Paulo speaks with Fernando, and Fernando occasionally asks Paulo to repeat some parts he did not understand.

Using \emph{AVoz} also gives Paulo an extra benefit that goes beyond intelligibility, and is related to \emph{voice identity} and the fact that he prefers to speak with his own voice rather than with a TTS system, as mentioned earlier in Section~\ref{subsec:how-Paulo-communicates}. With \emph{AVoz}, Paulo doesn't have to give up his own voice because it is present by default along with the corresponding TTS audio. In summary, Paulo gets the best of both worlds: voice identity and intelligibility. 
We quote two sentences from Weinberg et al.~\cite{Weinberg:2026} that could not be more appropriate: ``\emph{The value of a voice depends not only on intelligibility, but also on whether it enables users to represent themselves in ways that reflect their authentic identities.}" 
and also ``\emph{voice alignment runs deeper than accent or
language alone; it touches on belonging, self-recognition, and what
it means to be heard as who you are.}"

%
\section{Autonomy, Privacy, and Cost-Effectiveness}
\label{sec:ethics}

Customized ASR systems for people with speech disabilities sit at
the intersection of three key values in the field of
\emph{Computers \& Accessibility}, namely \emph{autonomy},
\emph{privacy}, and \emph{cost-effective assistive solutions}.
Advances in this field help enable those who have motor
impairments, in addition to their speech disabilities, to use
speech recognition-based interaction methods to control their
surrounding environment~\cite{Vertanen:2009,Hawley:2007}.
In this way, they can interact with home appliances,
computers, smartphones, and other physical artifacts, simply by 
``talking'' with them. 
Also, customized ASR systems for people with voice disabilities enables them to have a more
effective communication with others, enhancing their autonomy.

In general terms, this work shows that fine-tuning ASR models with 
speaker-specific training data is feasible with relatively low 
resources. In practice, a Mac Studio, a Mac mini, or a personal 
computer with a consumer-grade GPU is sufficient, yielding results 
of practical quality. This means that people with speech disorders, 
as well as their family members and caregivers, can fine-tune ASR 
models locally. This is, in practice, the key privacy guarantee, 
since no audio data ever leaves the user's machine and no speech 
is shared with third-party services~\cite{Condado:2023}.

Recorded speech is personal data, and in the case of a
person with a speech disorder, it can also reveal information
about their health. Under the General Data Protection Regulation,
data concerning health fall into a special category of personal
data whose processing is subject to stricter rules~\cite{Condado:2023}.
Moreover, voice recordings may be used to identify a person and,
when processed for this purpose, are considered biometric data,
which also belong to these special categories~\footnote{~\url{https://eur-lex.europa.eu/eli/reg/2016/679/oj}
(Accessed on Oct 12, 2026)}. As described by Nautsch et al.~\cite{Nautsch:2019},
speech carries much more information than its linguistic content,
including the speaker's identity, age, emotional state, and health,
which makes it a sensitive source of personal data.

In this specific work, all fine-tuning of Whisper models and all
inference run on local hardware. No audio recording ever leaves
the user's controlled machine. This is a \emph{privacy-by-design}
characteristic, consistent with the broader discussion of security
and privacy in assisted-living environments that we have
developed~\cite{Condado:2023,Condado:2022}, and also aligned with remarks made recently by Jacob Wobbrock at the end of his keynote speech at the 2026 ACM ASSETS conference. Addressing promises and pitfalls of AI for accessibility, Wobbrock raised concerns that to get the benefits of AI, people with disabilities are likely to have to abdicate more of their privacy than others, and he went on and mentioned that such is very unfortunate because people with disabilities are among those that could benefit the most from the technology~\cite{Wobbrock:2026}.

We couldn't agree more with Wobbrock's remarks, and our work exposes his concerns in a very clear way: by customizing an ASR locally, Paulo was able to get the benefits of AI without giving up his privacy. The same would not occur if the customization/training was done by an external entity on a cloud-based service. In our view this is an important advantage of our approach relative to other projects where people with disabilities have been participating as data contributors, and sometimes cannot even get the benefit of it as reported recently~\cite{Kobenova:2026}.

It is also worth mentioning  that access to AI is not the same for everyone. In the context of ASR, someone without a speech disability can use a system like Whisper right away without needing to customize it. Someone with a speech disability, however, needs to customize it for her/his own speech, and in doing so, losing privacy becomes a real risk unless the customization is performed locally as we advocate here.

A final dimension is the \emph{cost} of
implementing customized ASR systems for people with speech disorders.
LoRA fine-tuning of Whisper \emph{medium} can be performed on
relatively affordable hardware, and, with appropriate knowledge
and tools, makers should be perfectly able to develop
speech recognition-based AAC systems for themselves or members of
their community at a fraction of the cost of commercial AAC devices
or cloud-based services over a lifetime. This aligns with the
broader DIY-AT movement, in which the goal is not to design the
most sophisticated artifact, but to develop one that supports
people with disabilities in surpassing their functional
limitations and carrying out their daily activities~\cite{Parry-Hill:2017,Higgins:2024,Thorsen:2023,Lewis:2019}.

To ensure that anyone is capable of customizing their own model,
and to provide the information that maker communities need to reproduce the
fine-tuning pipeline end-to-end, we publish (i) the full list of
hyperparameters, 
(ii) the exact LoRA implementation used, and (iii) a step-by-step
protocol for recording, transcribing, splitting, and fine-tuning.
A person with a speech disorder, or a family member, caregiver, friend, 
or maker, should be able to
reproduce the pipeline within a couple days of GPU time
and a few hours of training data.


\section{Future Work}
\label{sec:future-work}

This work opens the door to a number of future research and development projects, some of which we describe below.

\begin{description}
	\item [Collect additional and more diverse data.]
	The customized ASR built for Paulo contains a relatively small dataset of audio samples. By collecting more data and performing further training, the accuracy of the system is likely to improve. Interestingly, we can use \emph{AVoz} itself to collect more data. Each time Paulo uses the application, the recorded audio and corresponding text transcription can be saved in our local server, and edited/reviewed later on to match the intended speech. Such editing/reviewing can obviously be done in the background, once in a while, not in real time. Since \emph{AVoz} is only making a few little mistakes, the number of corrections should not be large and can be made with little effort. Said differently, while a \emph{AVoz} is being used, it can collect more data as a side effect, effectively  bootstrapping an even better customized ASR. 
	
	Yet another advantage is that Paulo is likely to use \emph{AVoz} in  conversational and noisier environments, such as a lab, meeting, or even in the street, rather that reading pre-written sentences in a controlled environment. We expect that a subsequent fine-tuning with such data should make the system more robust. 
	Indeed, it  has been observed that in conversational situations where people with speech impairments use ASR systems to express an idea and communicate with others, the WERs are usually higher, both in unadapted and personalized models, and that the accuracy of these systems can be notably improved by including conversational speech samples in the training data~\cite{Tobin:2024}, something that can be easily accomplished by logging Paulo's interaction with \emph{AVoz}.
	
	\item [Address other languages.]
	Paulo's customized ASR works for the Portuguese language because it was based on audio samples spoken by him in Portuguese. A similar approach could be done for other languages, such as a English, a language he often needs to use to communicate with others. 
	Recording audio samples of him speaking pre-written sentences in English, however, does not seem to be an effective strategy. Not only it would be time-consuming, it would probably require a much larger training set to accomplish a similar accuracy level compared to Portuguese. Although Paulo understands English quite well, he has much more difficulty pronouncing it. When Paulo speaks in English, Fernando barely understands what he is trying to say: it is as if the severity of his speech disability suddenly increased by switching Portuguese to English.
	
	Fortunately a promising alternative exists. Whisper is a multi-task machine learning model being able to do speech recognition and speech translation. This means that we can bypass the need of fine-tuning Whisper for different languages --- we can use Whisper itself to translate the Portuguese text transcription to English (or any other language, for that matter) before sending it to the TTS system. Such an approach should work reasonably well, even though some errors that would be undetectable for Portuguese would definitely show up in English. (Recall that some of these undetectable errors in the context of a voice assistive application such as \emph{AVoz} result from oronyms or situations in which the sound of the text transcription is very similar to the intended sound. In the context of language translation such errors will not be handled in such a graceful manner.)
	
	\item [Voice-based text edition.]
	In addition to his speech disabilities, Paulo also has hand-motor disabilities due to his cerebral palsy condition, which makes him type much slower compared to most people. With a customized ASR enabled text editor he should be able to compose text much faster than doing so by typing. Even when occasional transcription mistakes occur, these can be corrected manually. We expect that such an editor would make the overall process of composing text, more effective and less tiring.
	
	\item [Control assisted living environments.]
	In the past we have been working on various aspects related to Assisted Living Environments (ALEs), from privacy and security issues to the design of appropriate user-interfaces for controlling them~\cite{Condado:2015,Condado:2022,Condado:2023}. The utilization of speech recognition for that endeavor has always been in the back of our minds, but never considered an option due to its low accuracy. With the present work, however, the ability for someone with a speech disability to control an ALE by issuing voice commands such as ``\emph{Turn on the light on the living room}" is now something within reach, not to mention that the vocabulary needed for controlling an ALE is a lot more restricted than the full vocabulary used in writing or in a regular conversation. 

	\item [Customized ASR for other people.]
	It remains to be seen if other people with speech disabilities can follow the approach that Paulo did, and build their own customized ASR system. Speech disabilities have a broad spectrum, ranging from minor disorders up to very severe ones. The success and amount of effort needed for the customization will certainly depend on the severity level of the speech disability, with the more severe cases requiring a larger amount of training data. Nonetheless, we hypothesize that if a person with a speech disability can be understood in an oral conversation by another person that is accustomed to talk to (as is the case of Paulo with respect to Fernando), then it should be possible to build a customized ASR to recognize that person's speech.

	\end{description}
	
\section{Summary and Conclusions}
\label{sec:conclusions}

By fine-tuning Whisper's \emph{medium} and
\emph{large-v3} models with LoRA on a few hours of his own speech, the
first author reduced the WER from 49\% to 17\% and from
30\% to 12\%, respectively, while the CER was reduced from 26\% to 7\% for \emph{medium}, and from 16\% to 5\% for \emph{large-v3}. The low CER values obtained indicate that the residual errors are minor phonemic substitutions of the kind
that a human listener can resolve from context. The resulting 
models generalize to utterances they never
heard before, confirming that they learned important patterns of
his speech rather than memorizing the training dataset. Building on
these models, the first author developed \emph{AVoz}, a mobile
application that converts his speech into synthesized voice in near
real-time, and without giving up his voice identity, restoring a form of oral communication that had been
unavailable to him.

The contribution of this work is methodological. We show
that the entire pipeline, from recording to fine-tuning to
inference, can be executed on consumer-grade hardware, that all data
remain under the user's control, and that the process can be
replicated by another person with speech disorders, perhaps with the help of a family
member or a friend. We publish a step-by-step training protocol, in the spirit of the DIY assistive technology movement.

Several limitations remain, and addressing them is central to our
future work. First, all training and test data was recorded in a
quiet office environment. In everyday use, however, the application
must operate in coffee shops, streets, and offices, where
background noise, reverberation, and competing voices degrade
recognition accuracy. We have observed that the model's
performance in such noisy settings is noticeably worse than in
quiet ones. 

Because Whisper’s models are fine-tuned for each individual user, 
generalization across speakers is neither expected nor required. What 
must be validated is whether the pipeline can be replicated by and for 
another person with a speech disorder, without requiring the technological 
expertise of the authors.

We finish the paper with some notes on learning.
Earlier in the text we mentioned that those accustomed to communicate orally with Paulo are
able to understand (most) of what he says, and it was that observation that gave us hope
that it should be possible to build an ASR system to recognize his speech. 

AI systems try to do tasks that require human intelligence, and the way they are built is
loosely inspired by the way the human brain works. Machine learning programs in
particular are able to learn from experience. This is what happened in the present work:
given examples of audio of sentences spoken by Paulo and the corresponding text of those sentences, the system learned important patterns of Paulo's speech, to the point that
it is able to generalize what it learned to recognize other audio samples spoken by Paulo that it never heard before.

But there is another lesson worth mentioning. 
If a machine, with appropriate training, is able to recognize Paulo's speech, so should a
human. The reality of our lived experience provides such evidence: Paulo's close friends and family are able to understand him because they are accustomed to talk with him --- their brains have
been trained to understand his way of speaking. What this reveals is that the reason why 
most people do not understand Paulo is because their brains have not been trained to
understand him. 
Unfortunately, or fortunately, machines are more patient than most people: they do not mind
spending time to listen and to try to understand a person with speech disabilities.

It remains to be seen, but we speculate that \emph{AVoz} itself can help people to better understand Paulo. By listening to his voice followed immediately by the corresponding TTS audio, people will hopefully learn to understand Paulo speaking with his own voice, to the point that eventually they will not need him to use \emph{AVoz} anymore, and will be able to  communicate directly with him the way his close friends and family members do. In other words, \emph{AVoz} can be used as a learning tool.
Interestingly this reveals the AI/ML learning paradigm in reverse. AI and ML programs learn from examples given to them as input. Likewise, people can learn to understand Paulo by being given examples of him speaking.

\section*{Acknowledgements}

Funda\c{c}\~ao para a Ci\^encia e a Tecnologia, I.P., Portugal, CENSE -- Center for
Environmental and Sustainability Research (UID/04085/2025 --
\url{https://doi.org/10.54499/UID/04085/2025}; UID/PRR/04085/2025 --
\url{https://doi.org/10.54499/UID/PRR/04085/2025}), and CHANGE -- Institute for Global
Change and Sustainability (LA/P/0121/2020 --
\url{https://doi.org/10.54499/LA/P/0121/2020}).

Paulo Condado's work was also sponsored by the Portuguese Foundation for Science and
Technology, FCT/MECI, under grant CEECIND/10962/2024.

\bibliographystyle{plain}
\bibliography{references}

\end{document}